\documentclass[aps,prl,twocolumn,superscriptaddress,showpacs,amsmath,amssymb,longbibliography]{revtex4-1}
\usepackage[T1]{fontenc}
\usepackage[utf8]{inputenc}
\usepackage{color}
\usepackage{babel}
\usepackage{amstext}
\usepackage{graphicx}
\usepackage{esint}
\usepackage{soul}
\usepackage[unicode=true,pdfusetitle,
 bookmarks=true,bookmarksnumbered=false,bookmarksopen=false,
 breaklinks=false,pdfborder={0 0 1},backref=false,colorlinks=false]
 {hyperref}
 \usepackage[normalem]{ulem}

\makeatletter
\@ifundefined{textcolor}{}
{%
 \definecolor{BLACK}{gray}{0}
 \definecolor{WHITE}{gray}{1}
 \definecolor{RED}{rgb}{1,0,0}
 \definecolor{GREEN}{rgb}{0,1,0}
 \definecolor{BLUE}{rgb}{0,0,1}
 \definecolor{CYAN}{cmyk}{1,0,0,0}
 \definecolor{MAGENTA}{cmyk}{0,1,0,0}
 \definecolor{YELLOW}{cmyk}{0,0,1,0}
}

\usepackage{multirow}

\makeatother

\begin{document}
\title{Universal semiclassical dynamics in disordered two-dimensional systems}

\author{\L ukasz Iwanek}
\affiliation{Institute of Theoretical Physics, Faculty of Fundamental Problems of Technology, Wroc\l aw University of Science and Technology, 50-370 Wroc\l aw, Poland}
\author{Marcin Mierzejewski}
\affiliation{Institute of Theoretical Physics, Faculty of Fundamental Problems of Technology, Wroc\l aw University of Science and Technology, 50-370 Wroc\l aw, Poland}
\author{Anatoli Polkovnikov}
\affiliation{Department of Physics, Boston University, 590 Commonwealth Avenue, Boston, Massachusetts 02215, USA}
\author{Dries Sels}
\affiliation{Department of Physics, New York University, New York, NY, USA}
\affiliation{Center for Computational Quantum Physics, Flatiron Institute, New York, NY, USA}
\author{Adam S. Sajna}
\affiliation{Institute of Theoretical Physics, Faculty of Fundamental Problems of Technology, Wroc\l aw University of Science and Technology, 50-370 Wroc\l aw, Poland}

\begin{abstract}
The dynamics of disordered two-dimensional systems is much less understood than the dynamics of disordered chains, mainly due to the lack of appropriate numerical methods. We demonstrate that a single-trajectory version of the fermionic truncated Wigner approximation (fTWA) gives unexpectedly accurate results for the dynamics of one-dimensional (1D) systems with moderate or strong disorder. Additionally, the computational complexity of calculations carried out within this approximation is small enough to enable studies of two-dimensional (2D) systems larger than standard fTWA. Using this method, we analyze the dynamics of interacting spinless fermions propagating on disordered 1D and 2D lattices. We find for both spatial dimensions that the imbalance exhibits a universal dependence on the rescaled time $t/\xi_W$, where in 2D the time-scale $\xi_W$ follows a stretched-exponential dependence on disorder strength.
\end{abstract}
\maketitle

\section{Introduction}
The dynamics of interacting and disordered quantum many-body systems have become the subject of numerous studies due to their peculiar anomalous behavior. This includes logarithmic-in-time relaxation processes observed in, e.g., entanglement entropy \cite{De_Chiara_2006,Znidaric_2008, Bardarson_2012,Serbyn_2013,Huang_2021} or correlation functions \cite{Mierzejewski_2016,Serbyn_2017,Sels_2021,Vidmar_2021}. Additionally, the fate of long-time dynamics in macroscopic disordered many-body systems has become a recent focus due to the stability problem of the many-body localized phase \cite{Panda_2019,Sierant_2020a,Sierant_2020b,Abanin_2021,Morningstar_2022,Suntajs_2020a,Suntajs_2020b,Sels_2022,Sels_2023, De_Roeck_2017}. Most studies on interacting and disordered quantum many-body systems have been conducted in 1D systems, leaving the 2D case largely unexplored. Theoretical exploration of 2D systems is challenging due to the exponential growth of the Hilbert space size for larger systems. Consequently, most of the contributions come from experiments in ultracold atomic systems \cite{Choi_2016,Schreiber_2015,Kondov_2015,Luschen_2017,Bordia_2016,Bordia_2017} and approximate numerical studies, such as Gutzwiller mean-field theory \cite{Hoi-Yin_Hui_2017}, tensor networks \cite{Wahl_2017,Wahl_2019,Chertkov_2021,Venn_2022}, the time-dependent variational principle \cite{Goto_2019,Paeckel_2019,Doggen_2020}, Hartree-Fock approximation \cite{Popperl_2021} and the truncated Wigner approximation (TWA) \cite{Davidson_2017,Schmitt_2019,Sajna_2020,Osterkorn_2022,Iwanek_2023,Rousse_2023}. Moreover, several works indicate that higher-dimensional systems exhibit many features similar to their 1D counterparts \cite{Sierant_2024,De_Roeck_2017,Urbanek_2018,Wiater_2018,Theveniaut_2020,Pietracaprina_2021,Venn_2022, Doggen_2020}. This suggests that the role of dimensionality should be studied more carefully.

We have recently demonstrated that the fermionic truncated Wigner approximation (fTWA) \cite{Iwanek_2023} provides valuable insights into the dynamics of disordered systems of interacting fermions. This semiclassical approach systematically overestimates the relaxation rates at strong disorder so that the resulting dynamics is faster than the dynamics obtained from the diagonalization of small systems.
While this property allows one to bind the actual quantum dynamics of 1D and 2D systems, fTWA fails to quantitatively describe the long-time expectation values of relevant observables.

In this work, we show that a single-trajectory version of fTWA is free from this drawback. For reasons that will be explained in the subsequent section, we call it a fermionic Gross-Pitaevskii (fGP) approach. We show that for moderate and strong disorder that results for imbalance obtained from fGP accurately follow results from the Lanczos propagation method up to the longest times and largest sizes accessible to the latter method. However, the computational complexity of fGP is relatively low (polynomial in the system size), enabling studies of much larger systems and longer evolution times. In particular, it allows for direct simulation of the experiments from Ref. \cite{Choi_2016}  in one half of a 2D system initially occupied by particles, whereas the other half is empty. Moreover, using fGP, we find for 1D and 2D systems that the imbalance exhibits a universal dependence on the rescaled time $t/\xi_W$, where the time-scale $\xi_W$ follows a stretched-exponential dependence on disorder strength. This universality qualitatively complies with the concept of the internal clock that has been previously formulated for 1D disordered chains \cite{Evers_2023}.

The paper is organized as follows. In Sec. \ref{ftwa}, the model and methods used in this manuscript are presented. In Sec. \ref{benchmark}, the validity of the single-trajectory version of fTWA, i.e., fGP method, is discussed, and its superior performance at intermediate and higher disorder compared to the standard fTWA method is justified in Sec. \ref{integrals}. Next, the universal dynamical behavior observed for the imbalance function is described in Sec. \ref{universality of the long-time dynamics}. At the end of the manuscript, the experimentally motivated studies in 2D systems are discussed (Sec. \ref{experimental}), and the paper concludes with Sec. \ref{summary}.

\section{Model and method} \label{ftwa}
In this paper, we investigate spinless fermions described by the Hamiltonian $\hat{H}=\hat{H}_{0}+\hat{H}_{V}$ with interaction and random potential on a lattice of $L$ sites with open boundary condition (OBC)
\begin{align}
\hat{H}_{0}&=-\frac{J}{2}\sum_{\left\langle ij\right\rangle }\left(\hat{c}_{i}^{\dagger}\hat{c}_{j}+h.c.\right)+\sum_{i}\epsilon_{i}\hat{n}_{i},\nonumber \\
\hat{H}_{V}&=V\sum_{\left\langle ij\right\rangle }\hat{n}_{i}\hat{n}_{j},\label{eq: hamiltonian}
\end{align}
where $\hat{c}_{i}^{\dagger}$ ($\hat{c}_{i}$) are the fermionic creation (annihilation) operators, $\hat{n}_{i}=\hat{c}_{i}^{\dagger}\hat{c}_{i}$ is the density operator on the site $i$ and $\langle ij\rangle$ denotes summation over nearest neighbor sites. The parameter $J$, representing hopping energy, will be used as an energy unit ($J=1$), $V$ is the nearest neighbor interaction strength and $\epsilon_{i}$ is the random potential uniformly distributed in the range $[-W,\,W]$.

\begin{figure*}[t]
\includegraphics[scale=1.09,page=1]{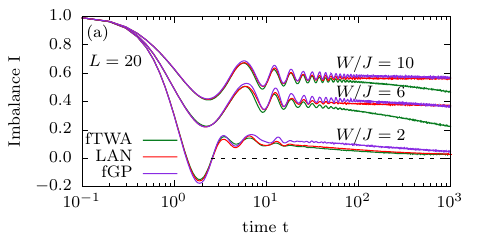}
\includegraphics[scale=1.09,page=1]{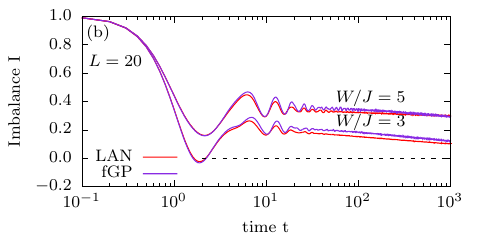}
\includegraphics[scale=1.09,page=1]{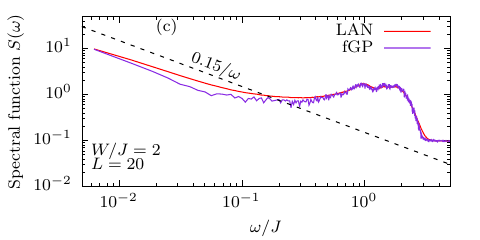}
\includegraphics[scale=1.09,page=1]{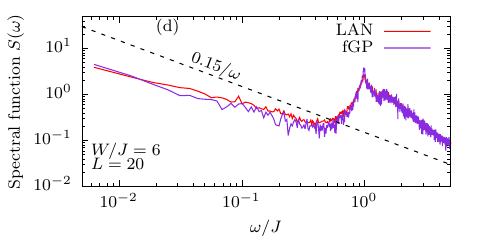}
\caption{Evolution in time for imbalance $I$ in panels (a) and (b) and spectral function $S(\omega)$ in panels (c) and (d) for various strengths of disorder. The system is a 1D chain of $L=20$ sites. Results for $\text{fTWA}$ (green line) and $\text{fGP}$ (violet line) are compared to the Lanczos method plotted as a (red line). Lanczos and $\text{fGP}$ are averaged over $1000$ disorder realizations. For $\text{fTWA}$ there are $500$ trajectories for each of $200$ disorder realization. The Hamiltonian parameters are set to $V=0.5$ and $J=1$.}
\label{fig: fTWA MF}
\end{figure*}

Semiclassical methods offer a unique combination of quantum operators expressed as a classical functions. We explore the dynamics of the spinless fermion Hamiltonian using the fermionic truncated Wigner approximation (fTWA). This approach is established in the Wigner-Weyl quantization procedure \cite{Davidson_2017}. The Hamiltonian $\hat{H}$ is mapped to its Weyl symbol $H_{W}$
\begin{equation}
	H_{W} =  -\frac{J}{2}\sum_{\left\langle ij\right\rangle }\left(\rho_{i j}+\rho_{ji}\right)+\sum_{i}(\epsilon_{i}\rho_{i i}+V\rho_{ii}^2)+ V\sum_{\langle ij\rangle}\rho_{i i}\rho_{j j}, \label{eq: H_W}
\end{equation}
where $\rho_{ij}$ are phase space variables obtained by mapping fermionic bilinears $\hat{E}^{i}_{j}=(\hat{c}_{i}^{\dagger}\hat{c}_{j}-\hat{c}_{j}\hat{c}_{i}^{\dagger})/2$. The phase space variables $\rho_{ij}$ satisfy the canonical Poisson bracket relations \cite{Davidson_2017, Polkovnikov_2010}. Local interaction term $V\rho_{ii}^2$ in Eq.~(\ref{eq: H_W}) improves 
an agreement between the approximate fTWA method and exact calculations. This term arises from the ambiguous procedure of quantization of fermionic operators \cite{Sajna_2020}. The dynamics of the phase space variables is described by the classical Hamilton equations of motion
\begin{equation} \label{Hamilton eqns}
	i\frac{d\rho_{kl}}{d t}  = \{\rho_{kl}, H_W\}= \sum_{m}  \left({\partial H_W\over \partial \rho_{lm} } \rho_{km}-{\partial H_W\over \partial \rho_{mk}} \rho_{ml}\right),
\end{equation}
The semiclassical expression for the expectation value of an observable $\hat{\mathcal{O}}(t)$ is given by the equation
\begin{equation}
\left\langle \hat{\mathcal{O}}(t)\right\rangle \approx\int \mathcal{O}_{W}[\rho_{kl}(t)] W(\{\rho^0_{kl}\})D\rho^0_{kl},\label{eq: average fTWA}
\end{equation}
where $\mathcal{O}_{W}$ is the Weyl symbol of $\hat{\mathcal{O}}$ and $D\rho^0_{kl}$ denotes integration over all independent phase space variables at the initial time. In this procedure, we express the Wigner function as a product of Gaussians
\begin{equation}
W(\{\rho_{kl}\})=\prod_{kl} \frac{1}{\sqrt{2\pi \sigma_{kl}^{2}}}\exp\left(-\frac{(\rho_{kl}-\mu_{kl})(\rho_{kl}^{*}-\mu_{kl})}{2\sigma_{kl}^{2}}\right),
\end{equation}
where $\mu_{kl}$ and $\sigma_{kl}$ are fixed by the initial exact expectation values. The first and second moments \cite{Davidson_2017} are defined as
\begin{align}
&\langle \hat{E}_{l}^{k}\rangle=\int\left[\prod_{mn}D\rho_{mn}\right]\rho_{kl}W(\{\rho_{mn}\}).\\
&\frac{1}{2} \langle  \hat{E}_{j}^{i}\hat{E}_{l}^{k} + \hat{E}_{l}^{k}\hat{E}_{j}^{i} \rangle=\int \left[\prod_{mn}D\rho_{mn}\right]\rho_{ij}\rho_{kl}W(\{\rho_{mn}\}).
\end{align}

In our study, besides the fTWA method, we also investigated dynamics in a case in which noise is neglected, i.e. $\sigma_{kl} \rightarrow 0$. The removal of noise in the initial condition results in one deterministic trajectory. This dynamic corresponds to solving noiseless Gross-Pitaevskii equations in bosonic systems \cite{Polkovnikov_2002}; therefore, we call it fermionic Gross-Pitaevskii (fGP) approach. Those equations are equivalent to the self-consistent Hartree-Fock approximation \cite{Popperl_2021,Rousse_2023}.

For short-time dynamics, one expects the fTWA to be more accurate than the fGP. This is due to the fact that fTWA is guaranteed to give asymptotically more accurate short time dynamics than a single trajectory fGP \cite{Anatoli_2003}. However, for longer times, this does not have to be the case. Indeed, in the strong disorder limit of the studied fermionic model, we find that fGP scenario is significantly more accurate than fTWA for the longer-time scales. We confirm this statement by carefully benchmarking both methods against exact Lanczos simulations that investigate relaxation dynamics.

It is worth noting that for the system size of length $L$, the computational complexity of both fGP and $\text{fTWA}$ scales like $L^{2}$, in contrast to the exact methods where full quantum Hilbert space scales as $2^{L}$. Moreover, fGP method, in comparison to fTWA, does not require sampling over initial noise, significantly reducing computational costs (this advantage of fGP will be discussed in the following sections). It is also worth adding that both fGP and fTWA are exact for all operators and non-equal time correlation functions in noninteracting systems and are nearly exact in the long-range interaction limit \cite{Sajna_2020}.

\section{Benchmark of fGP method}\label{benchmark}

To gain a better understanding and applicability of the semiclassical results, we benchmarked $\text{fGP}$ with the fTWA and Lanczos propagation method (LAN) in a one-dimensional (1D) system. The Lanczos method does not introduce systematic errors \cite{Park_1986, Mierzejewski_2010}, but it does impose a limitation on the system size, restricting it to $L\simeq 30$. The initial state is chosen as a charge density wave (CDW) state $|{010101...}\rangle$, where $|0\rangle$ represents an empty site and $|1\rangle$ is an occupied site.
In our comparison between methods, we used the imbalance function
\begin{equation}
I(t)=\frac{N_{o}(t)-N_{e}(t)}{N_{o}(t)+N_{e}(t)},
\label{eq: imbalance}
\end{equation}
with elements
\begin{equation}
N_{o/e}(t)=\sum_{i \in \text{initially (occupied/empty) sites}}\left\langle \hat{n}_{i}(t)\right\rangle.
\end{equation}
Moreover, in our relaxation dynamics analysis, we will utilize the spectral function of $I(t)$ defined as the Fourier transform 
\begin{equation}
S(\omega)=\int_{-\infty}^{\infty}I(t)e^{-i\omega t} dt=2\text{Re}\left[\int_{0}^{\infty}I(t)e^{-i\omega t}dt\right].
\label{sqo}
\end{equation}
In Fig.~\ref{fig: fTWA MF}(a), we present the time-dependent imbalance for three methods: $\text{fGP}$, $\text{fTWA}$ and $\text{Lanczos}$ obtained for a system with $L=20$ sites. Here, the limiting factor is the computational time of the Lanczos method. We examine three disorder strengths: small ($W/J=2$), intermediate ($W/J=6$) and strong ($W/J=10$). $\text{Lanczos}$ and $\text{fGP}$ results are averaged over $1000$ disorder realizations. The results of the fTWA (green line) are taken from Ref. \cite{Iwanek_2023} with $200$ disorder realizations.

For short times $t\lessapprox 20$, both semiclassical approaches accurately reproduce results from the Lanczos propagation method. In the case of the weak disorder ($W/J=2$), the accuracy of fTWA is preserved also for longer times, whereas $\text{fGP}$ slightly underestimates the decay rate of the imbalance. However, as disorder strength $W/J$ increases, $\text{fGP}$ begins to show good agreement with Lanczos results across all inspected time scales, while the decay rates obtained from fTWA are visibly overestimated. It is evident from Fig.~\ref{fig: fTWA MF}(a) that the $\text{fGP}$ significantly outperforms the standard semiclassical $\text{fTWA}$ method for intermediate and strong disorder. In Fig.~\ref{fig: fTWA MF}(b), we show results for other disorder strengths, $W/J=3$ and $W/J=5$, demonstrating that the consistency between fully quantum dynamics and the semiclassical $\text{fGP}$ is not limited to the regime of strong disorder. Unexpectedly, the simplest and most numerically undemanding approach ($\text{fGP}$) shows satisfactory agreement with the Lanczos propagation method over a broad range of disorder strengths.

In the case of a stronger disorder, the imbalance appears almost frozen and it is difficult to judge from 
Figs.\ref{fig: fTWA MF}(a) or \ref{fig: fTWA MF}(b) whether the slow but non-negligible dynamics of $I(t)$ is correctly captured by ($\text{fGP}$). Therefore, in figures \ref{fig: fTWA MF}(c) and \ref{fig: fTWA MF}(d), we present the spectral functions defined in Eq. (\ref{sqo}) obtained numerically from $\text{fGP}$ and the Lanczos propagation. In the case of weak disorder shown in panel (c), we observe a reasonable agreement between both methods, even though there are noticeable differences over a wide range of frequencies $\omega \sim 10^{-2} \div 10^{-1}$. In particular, both methods reproduce well-known behavior, $S(\omega) \propto 1/\omega$, which corresponds to almost logarithmic time-dependence of $I(t)$ \cite{Sels_2021, Mierzejewski_2016, Serbyn_2017, Vidmar_2021}.

\begin{figure}[t]
\includegraphics[scale=0.935,page=1]{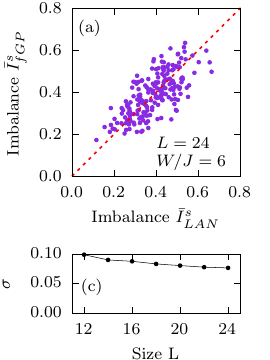}
\includegraphics[scale=0.935,page=1]{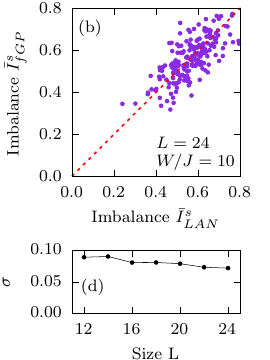}
\caption{Time averages of $\bar{I}^{s}_{\text{LAN}}$ and $\bar{I}^{s}_{\text{fGP}}$ represent single disorder imbalances for $200$ disorder realizations. Panel (a) corresponds to disorder strength of $W/J=6$, while panel (b) corresponds to $W/J=10$. The red dashed line represents the linear relation $\bar{I}^{s}_{\text{LAN}} = \bar{I}^{s}_{\text{fGP}}$. The initial state is a CDW with a system size of $L=24$ and the time average is in a time window of $t \in (0, 1000)$. Panels (c) and (d) show the deviation $\sigma$ for different sizes of the system for panels (a) and (b), respectively. The Hamiltonian parameters are set to $V=0.5$ and $J=1.0$. }
\label{fig: Single disorder}
\end{figure}

The bigger picture obtained from those results shows a significant advantage of $\text{fGP}$ over the standard $\text{fTWA}$, which holds true for a broad range of disorder strengths, except for the weak $W$. In Section V, we give an explanation for this observed improvement in $\text{fGP}$. A minor drawback of $\text{fGP}$ is excessive noise in the spectral function visible in Fig. \ref{fig: fTWA MF}(d). Analogous noise in fTWA is smaller due to averaging over multiple trajectories, which reflect averaging over initial quantum fluctuations. It can be easily addressed by computing more disorder realizations, which we avoid for comparison purposes, or by introducing smaller noise in initial conditions than required by Eq. (7). On the other hand, having only a single trajectory in the $\text{fGP}$ method reduces significantly computational time, allowing studies of quasi-2D systems, which we will discuss later in this paper.

Interestingly, we observe that the fTWA method exhibits a faster decay compared to the fGP method. One qualitative explanation is that, within fTWA, quantum fluctuations are represented by classical noise. When this noise accumulates (due to the dynamics) in one degree of freedom (such as a particular bilinear $\rho_{ij}$), the corresponding nonlinearity in the fTWA equations becomes very large (much larger than what is allowed by quantum mechanics) leading to faster relaxation \cite{Blakie_2008}. In contrast, such noise accumulation is not possible in fGP.

It is also worth noting that fGP behavior resembles simulations performed using the matrix-product-state-based TDVP \cite{Leviatan_2017}. This suggests that special care should be taken when the system possesses additional symmetries \cite{Wurtz_2020}.

\section{Single disorder realization analysis} \label{single disorder}
In the preceding section, we have discussed the time-dependence of the imbalance that was averaged over multiple disorder realizations. We demonstrated a very good agreement between results obtained from full quantum dynamics of disordered systems and a simple quasiclassical dynamics described by $\text{fGP}$. In this section, we show that this agreement does not arise as an artifact of averaging over various disorder strengths, but it holds true also for every disorder realization. To perform a quantitative comparison of results obtained for multiple realizations of disorder, we calculate a time-averaged imbalance 
\begin{equation}
\bar{I}_{\text{LAN/fGP}}^{s} = \frac{1}{\Delta t} \int_{0}^{\Delta t} I^{s} _{\text{LAN/fGP}}(t) dt, \label{timeav}
\end{equation}
where the index $s$ marks calculations of imbalance (see, Eq.~(\ref{eq: imbalance})) for a single disorder realization. We choose a large $\Delta t = 1000$ so that the time-averaged quantity in
Eq. (\ref{timeav}) corresponds to the imbalance in the long-time regime.

In Figs.~\ref{fig: Single disorder}(a) and ~\ref{fig: Single disorder}(b), we compare the time-averaged imbalances obtained from the fGP ($\bar{I}_{\text{fGP}}^{s}$) and Lanczos propagation 
($\bar{I}_{\text{LAN}}^{s}$). Each point shows a pair of results $(\bar{I}_{\text{LAN}}^{s}, 
\bar{I}_{\text{fGP}}^{s})$ obtained for the same realization of disorder.
Fig.~\ref{fig: Single disorder}(a) displays results for intermediate disorder ($W/J=6$), while Fig.~\ref{fig: Single disorder}(b) shows results for strong disorder ($W/J=10$). The system size is set to $L=24$. The points in Fig.~\ref{fig: Single disorder}(a) are clustered around smaller values of the imbalances than Fig.~\ref{fig: Single disorder}(b) in agreement with disorder-averaged results shown in Fig.~\ref{fig: fTWA MF}(a). Obviously, higher disorder strength $W/J$ preserves more information about the initial condition, resulting in higher values of imbalance. 

It is evident from these figures that fGP does not introduce significant systematic error (points are relatively evenly distributed around the diagonal that is set by the equality $\bar{I}_{\text{fGP}}^{s}=\bar{I}_{\text{LAN}}^{s}$, see red dashed lines in Fig. \ref{fig: Single disorder}. This result is consistent with our previous finding in Fig.~\ref{fig: fTWA MF}(a), where both methods were shown to exhibit good agreement in the disorder-averaged results. We also note that such an agreement does not hold true for the standard $\text{fTWA}$ \cite{Iwanek_2023}. Namely, one finds that $\text{fTWA}$ systematically underestimates the imbalance in the long-time regime. For this reason, results from $\text{fTWA}$ can be used only as a bound for the relaxation of imbalance, $I_{\text{LAN}}(t)\ge I_{\text{fTWA}}(t) \sim \log(t)$. 

The deviations between fGP and Lanczos methods are quantified in terms of 
\begin{equation}
\sigma^2 = \left\langle \left(\bar{I}_{\text{LAN}}^{s}-\bar{I}_{\text{fGP}}^{s}\right)^2 \right\rangle_{\text{dis}},
\label{eq: sigma}
\end{equation}
where $\langle ... \rangle_{\text{dis}}$ corresponds to averaging over 200 realizations of disorder. In panels ~\ref{fig: Single disorder}(c) and ~\ref{fig: Single disorder}(d), we observe that the deviation $\sigma$  weakly decreases with the system size. The quasiclassical approach is not expected to provide exact results, although better self-averaging might reduce 
the error as the system size increases. However, the simplicity of the numerical calculations carried out within $\text{fGP}$ allows one to study much larger systems and much longer times of evolution than the accurate quantum approaches. The results shown in the panels  ~\ref{fig: Single disorder}(c) and ~\ref{fig: Single disorder}(d) suggest that the errors generated in this case will not be larger than deviations observed for systems with $L \sim 20$.

\begin{figure}[t]
\includegraphics[scale=1.06,page=1]{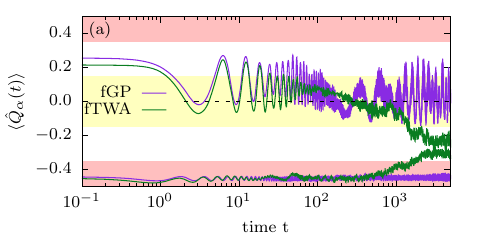}
\includegraphics[scale=1.06,page=1]{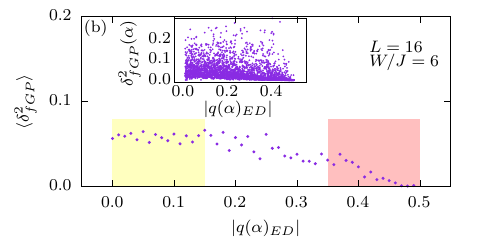}
\includegraphics[scale=1.06,page=1]{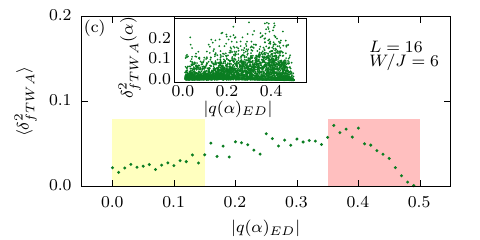}
\caption{In panel (a), there is an evolution of occupation $\langle  \hat{Q}_{\alpha} (t) \rangle$ for two sample cases. Examples of $\langle  \hat{Q}_{\alpha} (t) \rangle$ in the long-time regime approaches the red-shaded or the yellow-shaded areas are recognized as local regions with the strong or weak disorder. In the insets of panels (b) and (c), there are pairs of results: $[q(\alpha)_{ED},\delta_{fGP}^{2}(\alpha)]$ and $[q(\alpha)_{ED},\delta_{fTWA}^{2}(\alpha)]$ for $\alpha=1,...,L$ and $200$ realizations of disorder. The main panel in Figs. (b) and (c) show results for $q(\alpha)_{ED}$, which have been assigned to bins with  width $\Delta q_{\substack{\text{ED}}}=0.01$ and for each bin there is calculated the average deviation $\langle \delta_{fGP}^{2}  \rangle$ and $\langle \delta_{fTWA}^{2} \rangle$. In the $\text{fTWA}$ method, we calculate $400$ trajectories.}
\label{fig: Integrals}
\end{figure}

\section{Occupation of the single-particle Anderson states}  \label{integrals}
In this section, we discuss the origin of the counterintuitive observation that for moderate and strong disorder, the $\text{fGP}$ approach is more accurate than a more complex $\text{fTWA}$. We demonstrate that $\text{fGP}$ more accurately reproduces the dynamics of rare regions in the real space where disorder is particularly strong. Such spatially resolved information cannot be obtained from the imbalance. For this reason, we study the occupation of the single-particle Anderson states. To this end, we diagonalize the single-particle part of the Hamiltonian (\ref{eq: hamiltonian})
\begin{equation}
\hat{H_{0}}=\sum_{\alpha}\epsilon_{\alpha}\hat{Q}_{\alpha}+\text{const}.
\end{equation}
Here, $\epsilon_{\alpha}$ is the single-particle energy and 
$\hat{Q}_{\alpha}$ is the occupation of the Anderson state $|\alpha\rangle$
\begin{equation}
\hat{Q}_{\alpha}=\hat{a}^{\dagger}_{\alpha}\hat{a}_{\alpha}-\frac{1}{2},
\end{equation}
where $\hat{a}_{\alpha}^{\dagger}=\sum_{i}\langle i|\alpha \rangle\hat{a}_{i}$.   
In order to analyze the long-time behavior of various $\hat{Q}_{\alpha}$ for different realizations of disorder, we calculate the time-averaged expectation values
\begin{equation}
q(\alpha) = \frac{1}{\Delta t} \int_{0}^{\Delta t} {\rm d}t \langle  \hat{Q}_{\alpha} (t) \rangle,
\end{equation}
for $\Delta t=2000$ using exact diagonalization (ED) as well as $\text{fGP}$ and $\text{fTWA}$.
Local regions with strong disorder correspond to sites $i$ where $|\langle i|\alpha \rangle|$ 
is large and $|q_{\alpha}| \simeq 1/2$, whereas regions with weak disorder correspond to large $|\langle i|\alpha \rangle|$ and small $|q_{\alpha}| \simeq 0$. In Fig.~\ref{fig: Integrals}(a), we show two sample cases for $\langle  \hat{Q}_{\alpha} (t) \rangle$. Local regions with strong or weak disorder are identified as cases in which $\langle  \hat{Q}_{\alpha} (t) \rangle$ in the long-time regime approaches the red-shaded or the yellow-shaded areas, respectively.

The errors introduced by the semiclassical methods are quantified in terms of the mean square deviations  
\begin{equation}
\delta_{m}^{2}(\alpha)= \frac{1}{\Delta t} \int_{0}^{\Delta t} {\rm d}t
\left[ \langle  \hat{Q}_{\alpha} (t) \rangle_{\rm ED} - \langle  \hat{Q}_{\alpha} (t) \rangle_m \right]^{2},
\end{equation}
where $\langle  \hat{Q}_{\alpha} (t) \rangle_{\rm ED}$ is obtained from ED, whereas the index $m$ refers to the semiclassical method, i.e., either $m=\text{fGP}$ or $m=\text{fTWA}$. Insets in figures
\ref{fig: Integrals}(b) and  \ref{fig: Integrals}(c) show, respectively, pairs of results 
$[q(\alpha)_{ED},\delta_{fGP}^{2}(\alpha)]$ and $[q(\alpha)_{ED},\delta_{fTWA}^{2}(\alpha)]$ for 
$\alpha=1,...,L$ and $200$ realizations of disorder. Due to the limitations of ED, these results have been obtained for $L=16$. The main panel in Figs. \ref{fig: Integrals}(b) and  \ref{fig: Integrals}(c) show coarse-graining analysis of points shown in the insets. To this end, numerical results for $q(\alpha)_{ED}$ have been assigned to bins of the width $\Delta q_{\substack{\text{ED}}}=0.01$ and for each bin, we have calculated the average deviation $\langle \delta_{fGP}^{2}  \rangle$ and $\langle \delta_{fTWA}^{2} \rangle$. Consequently, results in Figs. \ref{fig: Integrals}(b) and  \ref{fig: Integrals}(c) show correlations between the long-time occupations of the Anderson states and the errors in both semiclassical approaches. One observes that fTWA is more accurate than fGP for small $|q(\alpha)|$, which correspond to weakly disordered regions (yellow-shaded areas). In contrast, fGP gives more accurate results for strongly disordered regions with $|q(\alpha)|\simeq 0.5$ (red-shaded areas). This demonstrates that the advantage of fGP over fTWA is not universal. This advantage occurs for systems with moderate and strong disorder, where fTWA overestimates the decay rates. 

\begin{figure*}[t]
\includegraphics[scale=1.09,page=1]{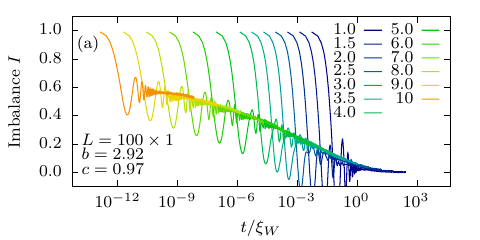}
\includegraphics[scale=1.09,page=1]{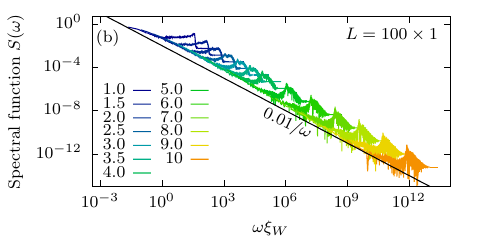}
\includegraphics[scale=1.09,page=1]{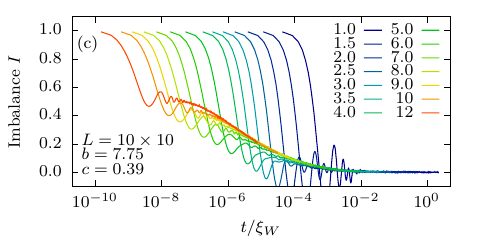}
\includegraphics[scale=1.09,page=1]{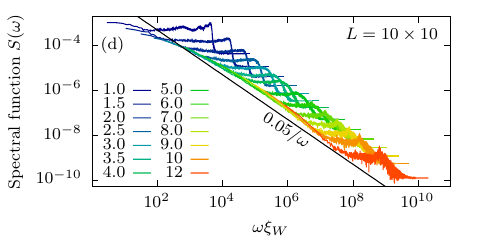}
\caption{Panels (a) and (c) show universal behavior of imbalance $I$ as a function of rescaled time $t/\xi_{W}$ with the exponential function $\exp(bW^{c})$ for both 1D and 2D. Panels (b) and (d) show the spectral function $S(\omega)$ obtained for a wide range of the disorder strength $W/J \in [1,12]$. System sizes for 1D and 2D are chosen with geometries of $L=100\times1$ and $L=10\times10$. The Hamiltonian parameters are set to $V=0.5$ and $J=1$. There are 200 disorder realizations for each disorder strength.}
\label{fig: Universality}
\end{figure*}

\section{Universal dynamics of imbalance}  \label{universality of the long-time dynamics}
In the 2D systems, we consider the initial product state in the form of CDW stripes. Those structures are directly accessible in experiments \cite{Bordia_2016,Bordia_2017}. Using the advantage of the $\text{fGP}$ method in simulating larger systems, we show that relaxation dynamics in disordered 2D systems exhibit universal behavior. More precisely, we demonstrate that the results in Fig.~\ref{fig: fTWA MF} obtained for various strengths of disorder, $I_W(t)$, overlap when plotted against a rescaled time, $I_W(t/\xi_W)$. Interestingly, we show that such behavior is even more pronounced than in 1D systems \cite{PhD_Wurtz_2020}.

For each disorder strength, $W$, we generate a set of times, $t_i$ and the corresponding (disorder-averaged) imbalances, $I_W(t_i)$. We look for a scaling function $\xi_W$ such that the set of points 
\begin{equation}
{\cal S}=\left\{ \left[ \frac{t_{i}}{\xi_{W}},I_W\left( \frac{t_{i}}{\xi_{W}} \right) \right] \right\},
\label{ep points}
\end{equation}
obtained for different $W$ collapses on a single smooth line. Motivated by the fact that the transport coefficients of disordered systems decay exponentially with $W$ \cite{Prelovsek_2021,Prelovsek_2023,Nandy_2024}, we have tested the following stretched-exponential form of the scaling function  
\begin{equation}
 \xi_{W}=\exp(bW^{c}).
 \label{eq: rescaled_time}
\end{equation}
It involves two fitting parameters, $b$ and $c$, which are determined via a stochastic algorithm from \cite{Storn_1997}. First, the set of points from Eq.~(\ref{ep points}) is sorted in ascending order according to the rescalled times, $t_i/\xi_W$. For convenience, we label the sorted points with the index $\alpha$
\begin{align}
\tau_{\alpha}&=\frac{t_{i}}{\xi_{W}}, \quad \tau_{\alpha} \le\tau_{\alpha+1},  \\
I_{\alpha}&=I_{W}\left( \frac{t_{i}}{\xi_{W}} \right).
\end{align}
The fitting parameters are obtained by minimizing the cost function
\begin{equation}
C_{I}=\sum_{\alpha} \left( I_{\alpha+1}-I_{\alpha} \right)^{4}.
\end{equation}
We note from Fig. \ref{fig: fTWA MF} that the imbalance exhibits numerous oscillations at the beginning of the evolution, where universal behavior is not expected. To address this feature, for each $W$ we used $N=49000$ equally spaced times $10^2 \le t_i \le 5\times10^4 $. In the fitting procedure, we used imbalances obtained for $W \in \{1,1.5,2,2.5,\ldots,4,5,6,\ldots,10\}$ for 1D and $W \in \{1,1.5,2,2.5,\ldots,4,5,6,\ldots,12\}$  for 2D. 

In Figs.~\ref{fig: Universality}(a) and ~\ref{fig: Universality}(c), we present the imbalance function $I$ with exponentially rescaled time $t/\xi_{W}$. Figs. ~\ref{fig: Universality}(b) and ~\ref{fig: Universality}(d) consist of the spectral function after the fitting procedure. We study two system geometries: 1D with a size of $L=100 \times 1$ (Figs. ~\ref{fig: Universality}(a) and ~\ref{fig: Universality}(b)) and 2D with a size of $L=10 \times 10$ (Figs. ~\ref{fig: Universality}(c) and ~\ref{fig: Universality}(d)). The presented results of the $\text{fGP}$ are averaged over 200 disorder realizations.

In Fig. ~\ref{fig: Universality}(a), we show the universal behavior of imbalance in a wide range of disorder strengths. Challenges inherent to the intricate nature of 1D systems manifest in the analysis presented in Fig. 4(a), where we show the universal behavior of imbalance across a wide range of disorder strengths. Overlapping parts in disorder range of W/J = [5; 10] show less smooth alignment compared to weaker disorder strength. This problem could be potentially addressed by increasing the number of disorder realizations and extending the time evolution in our calculations. However, the exponential scaling makes it difficult to achieve, even for the fGP method.

The spectral behavior in Fig. ~\ref{fig: Universality}(b) shows a clear $1/\omega$ trend, reflecting a universal behavior in 1D disordered lattice systems. This result extends previous work on 1D systems \cite{Sels_2021, Mierzejewski_2016, Serbyn_2017, Vidmar_2021} showing that $1/\omega$ behavior has a universal character and is not tied to quantum mechanics. 

For the 2D system, we observe excellent accuracy of the fit across a wide range of disorder strengths (see, Fig. \ref{fig: Universality}(c)). Results from both weak and strong disorder align much more smoothly with the overall character when compared to the 1D case. The improvement of universal behavior for 2D systems suggests that higher dimensional lattices, where quantum fluctuations are expected to play a weaker role, follow universal behavior more easily.

Fig. ~\ref{fig: Universality}(d) demonstrates good alignment of the spectral function in a wide range of disorder strengths, except for $W/J=[1,2]$. The $1/\omega$ trend is also clearly visible in the 2D case. The extra degrees of freedom make semiclassical results less noisy.

\section{Experimentally motivated framework} \label{experimental}
In this section, we will leverage $\text{fGP}$ ability to simulate larger systems in order to create a setup closely related to cold-atom systems (we follow the idea from Ref. \cite{Choi_2016}).  First, we modify the initial state from half-filling to quarter-filling, where we divide the system into two parts (left and right). Each fermion is placed randomly in the left part, while the right one is unoccupied. The idea behind quarter-filling is to avoid creating a fully occupied block on the left side and leave space for particles to move.

To quantify the asymmetry in the initial density, we used the imbalance function
\begin{equation}
K(t)=\frac{N_{L}(t)-N_{R}(t)}{N_{L}(t)+N_{R}(t)},
\label{eq: imbalanceK}
\end{equation}
with elements
\begin{equation}
N_{L/R}(t)=\sum_{i \in \text{(LEFT/RIGHT) side of the lattice}}\left\langle \hat{n}_{i}(t)\right\rangle.
\end{equation}

\begin{figure}[t]
\includegraphics[scale=1.06,page=1]{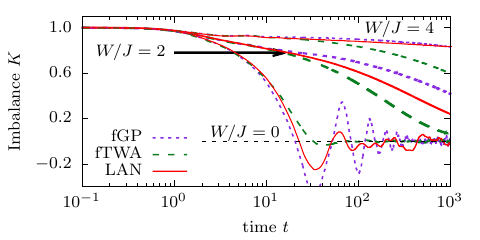}
\caption{Time evolution of the imbalance $K$ for various disorder strengths $W/J$. The geometry is a 1D system of $L=24$ sites with $6$ fermions placed randomly on the left side. There are $20$ different random states and each disorder line is averaged over $200$ disorder realizations. In fTWA, we calculated $200$ trajectories. The Hamiltonian parameters are set to $V=0.5$ and $J=1$.}
\label{fig: Experiment}
\end{figure}

First, we will check the applicability of semiclassical results in the quarter-filling framework. In Fig.~(\ref{fig: Experiment}), we test $\text{fGP}$ with the fTWA and Lanczos method in a 1D system with $L=24$ sites. There are $20$ random states with $6$ fermions randomly placed on the left side of the lattice. We choose three disorder strengths: $W/J=0, 2, 4$.

Starting with the case of $W/J=0$, we can see that all methods qualitatively reproduce the $K \rightarrow 0$ at longer times. The $\text{fTWA}$ method is able to catch relaxation faster than $\text{fGP}$
but fails to reproduce oscillations at early times. On the other hand, the $\text{fGP}$ oscillations have an overestimated magnitude.

At disorder $W/J=2$, there is an interesting relationship between the three methods. Lanczos propagation is in the middle, while $\text{fTWA}$ and $\text{fGP}$ are setting the upper and lower bounds. The behavior of slower decay of imbalance in $\text{fGP}$ looks similar to previously observed in Fig. ~\ref{fig: fTWA MF}(a) for a half-filling framework.

Results for disorder strength W/J=4 show that the imbalance from both the Lanczos propagation and the fGP exhibits similarly slow behavior. This behavior aligns with Figs.~\ref{fig: fTWA MF}(a) and ~\ref{fig: fTWA MF}(b), where $\text{fGP}$ correctly reproduces results for intermediate and strong disorder, thus strengthening method consistency. The faster decay of imbalance in $\text{fTWA}$ also coincides with our previous findings.

To fully utilize the strength of $\text{fGP}$, we investigate a disordered 2D fermionic lattice of $L=10\times10$. We implement the framework of quarter-filling with $25$ fermions placed randomly on the left side of a square lattice.
\begin{figure}[t]
\includegraphics[scale=1.06,page=1]{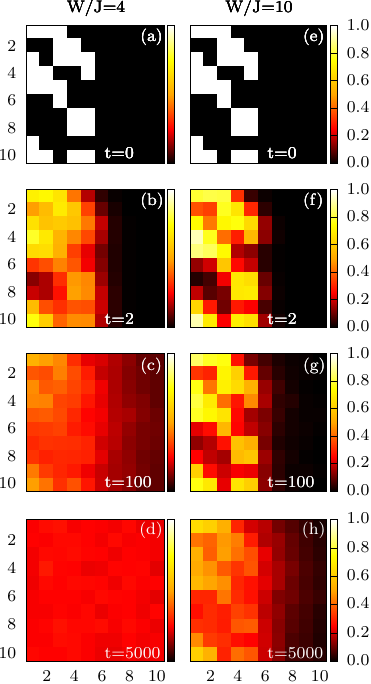}
\caption{Evolution of the initial state for two disorder strengths: $W/J=4$ and $W/J=10$. The geometry of the model is a 2D system of $L=10\times10$ with $25$ fermions placed randomly on the left side. We show four snapshots of the density evolution: (a) and (e) for time $t=0$, (b) and (f) for time $t=2$, (c) and (g) for time $t=100$, (d) and (h) for time $t=5000$. The Hamiltonian parameters are set to $V=0.5$ and $J=1$. There are $50$ disorder realizations for each disorder strength.}
\label{fig: Experiment2}
\end{figure}
The time evolution of initial density for two disorder strengths: $W/J=4$ (a-d) and $W/J=10$ (e-h) is shown in the panels of Fig.~(\ref{fig: Experiment2}). Snapshots consist of lattice density at four different times. Panels of Figs. ~\ref{fig: Experiment2}(a) and ~\ref{fig: Experiment2}(e) demonstrate the initial state at the beginning of the evolution at time $t=0$. There are visible islands of randomly placed fermions marked by a white color. As the system begins its evolution, we can distinguish two mechanisms. The first one will be responsible for transport between occupied and unoccupied sites on the left side of the lattice. The second will be responsible for transport between the left and right side of the system. 

At short time scales of $t=2$, Figs. ~\ref{fig: Experiment2}(b) and ~\ref{fig: Experiment2}(f), we can observe that the first process has a stronger impact. The initial density starts spreading across the left side. This behavior is more pronounced at smaller disorder of $W/J=4$. By the time the density has fully relaxed in the vertical direction, the second process has only just begun to manifest itself in both disorder cases.

Panels of Figs. ~\ref{fig: Experiment2}(c) and ~\ref{fig: Experiment2}(g) show intermediate times of $t=100$, where we can distinguish the high influence of disorder on the system's evolution. In the weaker disorder case $W/J=4$, the left side density spreads into a wave that begins flowing into the right side of the system. For the $W/J=10$ disorder strength, the first process continues on the left side, while the second process has minimal impact. 

The long-term behavior at $t=5000$ is represented in Figs. ~\ref{fig: Experiment2}(d) and ~\ref{fig: Experiment2}(h). At disorder strength $W/J=4$ system completely loses information about the initial state, which is visible as a uniformly spread probability of $\hat{n}_{i}=0.25$ for a given site $i$. On the other hand, at strong disorder $W/J=10$ with the help of the first process, the density spreads on the left side and starts merging into the wave flowing on the right side. Comparing Figs.~\ref{fig: Experiment2}(c) and ~\ref{fig: Experiment2}(h) show the influence of disorder on slowing down the dynamics of the system. Increasing the disorder strength from $W/J=4$ to $W/J=10$ leads to slowing down dynamics by an order of $\mathcal{O}(10^{2})$. This finding strengthens the proposed exponential scaling in Eq.~(\ref{eq: rescaled_time}).

\begin{figure}[t]
\includegraphics[scale=1.06,page=1]{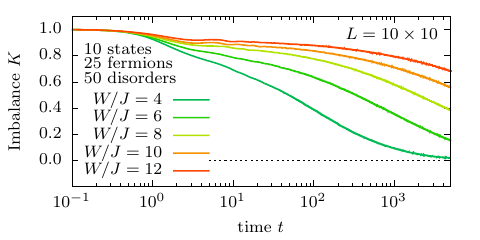}
\caption{Evolution in time for imbalance $K$ for various disorder strengths $W/J$. The geometry is a 2D system of $L=10\times10$ sites with $25$ fermions placed randomly on the left side. There are $10$ different random states and each disorder line is averaged over $50$ disorder realizations. The Hamiltonian parameters are set to $V=0.5$ and $J=1$.}
\label{fig: Experiment3}
\end{figure}

While Fig.~(\ref{fig: Experiment2}) shows site-resolved images that present the behavior of the system from the level of individual sites, Fig.~(\ref{fig: Experiment3}) illustrates imbalance $K$ results averaged over the left and right sides of the system, treating them as two distinct units.
The imbalance $K$ (Eq.~(\ref{eq: imbalanceK})) is computed for 2D geometry of $L=10\times 10$ with $10$ different random states (only the left side is occupied). There are five different disorder strengths: $W/J=4, 6, 8, 10, 12$ and each line is averaged over $50$ disorder realizations. In Fig.~(\ref{fig: Experiment3}), the weakest disorder line of $W/J=4$ shows that at the end of the examined time evolution, there is an imbalance $K \simeq 0$, suggesting that the system has no information left about initial state and probability has spread evenly, consistent with Fig.~\ref{fig: Experiment2}(d). Stronger disorder lines of $W/J>4$ have a clear descending trend, indicating that for sufficiently long time scales, the initial information should also vanish. This late-time behavior is difficult to properly address due to long computational times because the time scale varies with disorder strength as $\xi_W$ (see discussion in Sec. \ref{universality of the long-time dynamics}).

The conclusions from Fig.~(\ref{fig: Experiment}) are very similar to those previously drawn from the half-filling framework in Fig.~\ref{fig: fTWA MF}. The $\text{fGP}$ works better at intermediate disorder, while $\text{fTWA}$ works better at weak disorder. Extending simulations to 2D systems in both Fig.~(\ref{fig: Experiment2}) and Fig.~(\ref{fig: Experiment3}) show the importance of examining longer time evolution. Increased disorder slows down dynamics; compare Figs.~\ref{fig: Experiment2}(c) and ~\ref{fig: Experiment2}(h). This indicates that in both numerical and experimental setups, in order to understand the long-time behavior of such systems, it is important to increase time scales or carefully study scaling collapse, as shown in Fig. ~\ref{fig: Universality}.

\begin{figure}[t]
\includegraphics[scale=1.06,page=1]{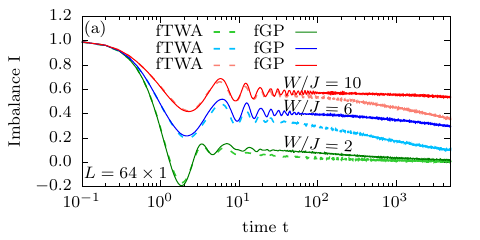}
\includegraphics[scale=1.06,page=1]{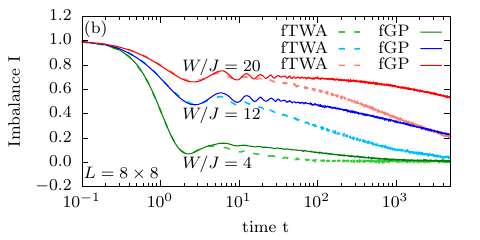}
\caption{Time evolution of the imbalance $I$ for various disorder strengths $W/J$. Two geometries are considered: a 1D system of $L=64\times1$ with CDW and a 2D system of $L=8\times8$ with stripes CDW initial state. For fGP there are $200$ disorder realizations, while for fTWA there are $30$ trajectories for each of $50$ disorder realizations. The Hamiltonian parameters are set to $V=0.5$ and $J=1$.}
\label{fig: initialnoise1d2d}
\end{figure}

At the end of this section, it is also important to discuss how the effect of initial quantum noise on the system's dynamics depends on its dimensionality. It turns out that quantum noise causes a stronger destabilization of memory effects in a 2D system than in the 1D case (see Fig. \ref{fig: initialnoise1d2d}). This observation aligns with the qualitative explanation provided in Sec. \ref{benchmark}. Specifically, a larger phase space in the 2D case leads to greater noise accumulation in the fTWA method compared to the fGP approach, which is noise-free. Consequently, the nonlinearities present in the Hamilton equations (Eq. \ref{Hamilton eqns}) result in a faster decay of imbalance functions. Additionally, we find that the spectral functions calculated from the data in Fig. \ref{fig: initialnoise1d2d} do not show qualitative differences between the fGP and fTWA theories, further suggesting that the $1/\omega$ noise has a classical origin.

\section{Conclusions} \label{summary}
We have studied the dynamics of spinless fermions in disordered systems using a semiclassical approach based on the fermionic truncated Wigner approximation (fTWA). We have focused on its simplified version ($\text{fGP}$) which accounts only for a  single trajectory dynamics. First, we have benchmarked $\text{fGP}$ against $\text{fTWA}$.
To this end, we have compared results for imbalance obtained using fGP and fTWA with accurate numerical data obtained from the Lanczos propagation method for one-dimensional systems. This comparison shows, quite unexpectedly, that   $\text{fGP}$  significantly outperforms the semiclassical $\text{fTWA}$ in capturing the dynamics in the regimes of intermediate to strong disorder. In contrast to fTWA, fGP does not introduce systematic errors and accurately reproduces both the magnitude of the imbalance as well as its asymptotic time-variation. The fact that the fGP describes the dynamics of the studied systems so well indicates that this ultraslow dynamics is not specific only to quantum systems, as it can be well reproduced by classical dynamics.

The simplicity of numerical calculations based on the fGP allowed us to study both 1D and 2D systems. For both cases, the fGP method shows a $1/\omega$ behavior of the spectral function, known as $1/f$ noise indicating slow, logarithmic-in-time, relaxation. We have also demonstrated a universal dependence of imbalance on the rescaled time $t/\xi_W$, where the time-scale $\xi_W$ follows a stretched-exponential dependence $\xi_{W}=\exp(bW^{c})$ with $c\simeq 1$ and  $c\simeq 1/2$ for 1D and 2D systems, respectively. Except for the latter difference, 1D and 2D systems show overall similar dynamics that, for sufficiently strong disorder, appear almost frozen within any finite time-window. However, at least the 2D systems should eventually thermalize \cite{De_Roeck_2017}.

Finally, by simulating relaxation processes similar to those in experimental setups conducted with ultracold atomic gasses \cite{Choi_2016}, we have shown that, in comparison to the standard fTWA, the fGP method provides highly reliable results under stronger disorder. Moreover, it turns out that the fGP and fTWA methods serve as powerful tools for estimating the lowest and highest possible relaxation rates, which can be valuable, for example, in designing future experimental setups.

\section{Acknowledgments} \label{acknowledgments}
A.S.S. would like to thank Jonathan Wurtz for valuable discussions. A.P. acknowledges support from NSF under Grant No. DMR-2103658 and the AFOSR under Grant No. FA9550-21-1-0342. Ł.I. and M.M. acknowledge support by the National Science Centre, Poland via Project No. 2020/37/B/ST3/00020. Numerical studies in this work have been carried out using resources provided by the Wroclaw Centre for Networking and Supercomputing.  \footnote{http://wcss.pl}, Grant No. 551 (1692966935) and Grant No. 1721138190.

\bibliographystyle{apsrev4-1}
\bibliography{library.bib}

\end{document}